\documentclass[a4paper,notoc,hyper]{JHEP3} 

\usepackage{epsfig,multicol,bbm}

\title{Determination of supersymmetric masses using kinematic fits at the LHC}

\author{Christian Autermann, Benedikt Mura, Christian Sander, Hannes Schettler, Peter Schleper \\
	Hamburg University, Luruper Chaussee 149, 22761 Hamburg, Germany\\
	E-mail: \email{christian.sander@desy.de}}

\preprint{hep-ph/0911.2607}

\abstract{
In case of the discovery of supersymmetry at the LHC, the goal will be to identify the underlying theory, its fundamental parameters, and the masses of SUSY particles. We followed here the approach to reconstruct the decay chains in SUSY events under the assumption of common intermediate masses. These masses cannot be extracted from each event because of the unmeasured LSP momenta in case of $R$-parity conservation. But an ensemble of events can be over-constrained, if the decay chains are long enough, such that enough mass constraints are available. Here, we present a new method combining a) a SUSY mass scan, b) a kinematic fitting based on a genetic algorithm for decay chain reconstruction, and c) the usage of angular decay information to suppress the background from other SUSY processes. Taking into account the full combinatorial background and experimental resolutions in the most difficult case of the fully hadronic decay mode, we demonstrate, within one SUSY scenario, that this method can be used to derive a probability map of the SUSY parameter space.
}

\keywords{Supersymmetry, phenomenology, kinematic fits, parameter determination, LHC}

\begin{document} 

\section{Introduction}\label{Introduction}

One possible scenario at the Large Hadron Collider (LHC) is the observation of a significant number of events above the Standard Model (SM) background with many jets and/or leptons in the final state, as well as large amounts of missing transverse energy $\not{\hspace{-3pt}E}_T$. Even if the cross sections of these processes are large, as predicted by some models, the identification of the underlying theory is rather challenging. The determination of the masses of the involved new particles is not straight forward, if one or more particles escape the detector unseen, as it occurs in $R$-parity conserving supersymmetric (SUSY) extensions of the SM \cite{Martin:1997ns}, where the lightest supersymmetric particle (LSP) is in many scenarios only weakly interacting. A further major challenge is the suppression of the combinatorial background, which can be of order $10^{3}$, if a large number of jets are present.

There are various approaches to mass determination in SUSY events, which can be divided into two major groups: 
\begin{itemize}
  \item[(a)] Determination of endpoints in invariant mass or other kinematic distributions. These observables are typically sensitive to mass differences, and by the combination of a few of them, the masses of the involved particles can be accessed \cite{Hinchliffe:1996iu}-\cite{Han:2009ss}.
  \item[(b)] Reconstruction of entire decay chains on event-by-event basis. This is only possible, if there are at least as many constraints as unknowns for the problem \cite{Nojiri:2007pq}-\cite{Webber:2009vm}.
\end{itemize}
Here, we follow the second approach and, for the first time, take into account the experimental uncertainty of the measurements for both the event simulation and, by means of a kinematic fit, in the event reconstruction \cite{Schettler:2009, SanderSUSY09}. In order to develop a method which is able to deal with high combinatorial background, we demonstrate our method on the fully hadronic decay mode of $\tilde{q}\tilde{g}$ production with seven jets in the final state (see Fig.~\ref{fullhad}).

\DOUBLEFIGURE[b]{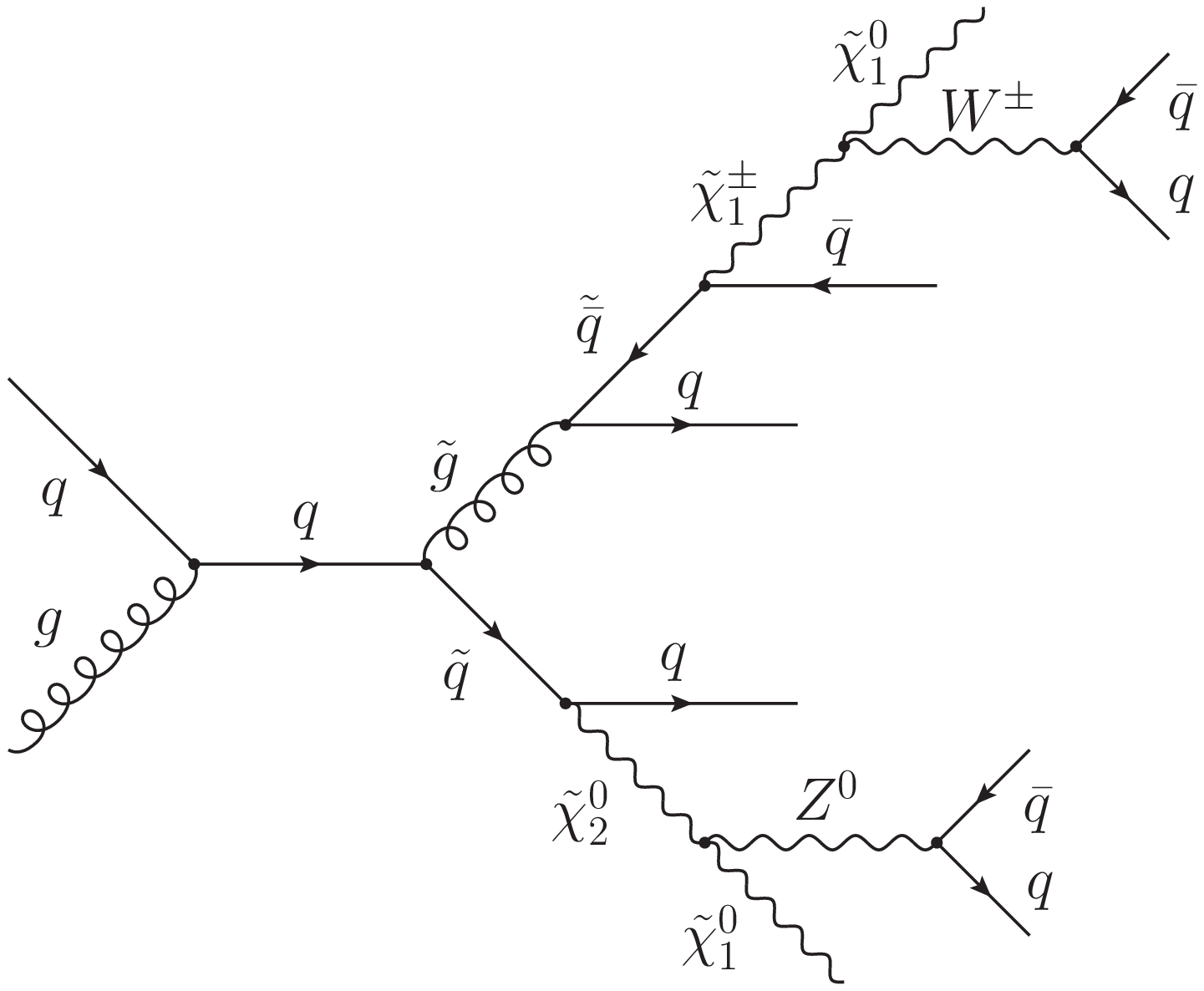,width=0.49\textwidth}{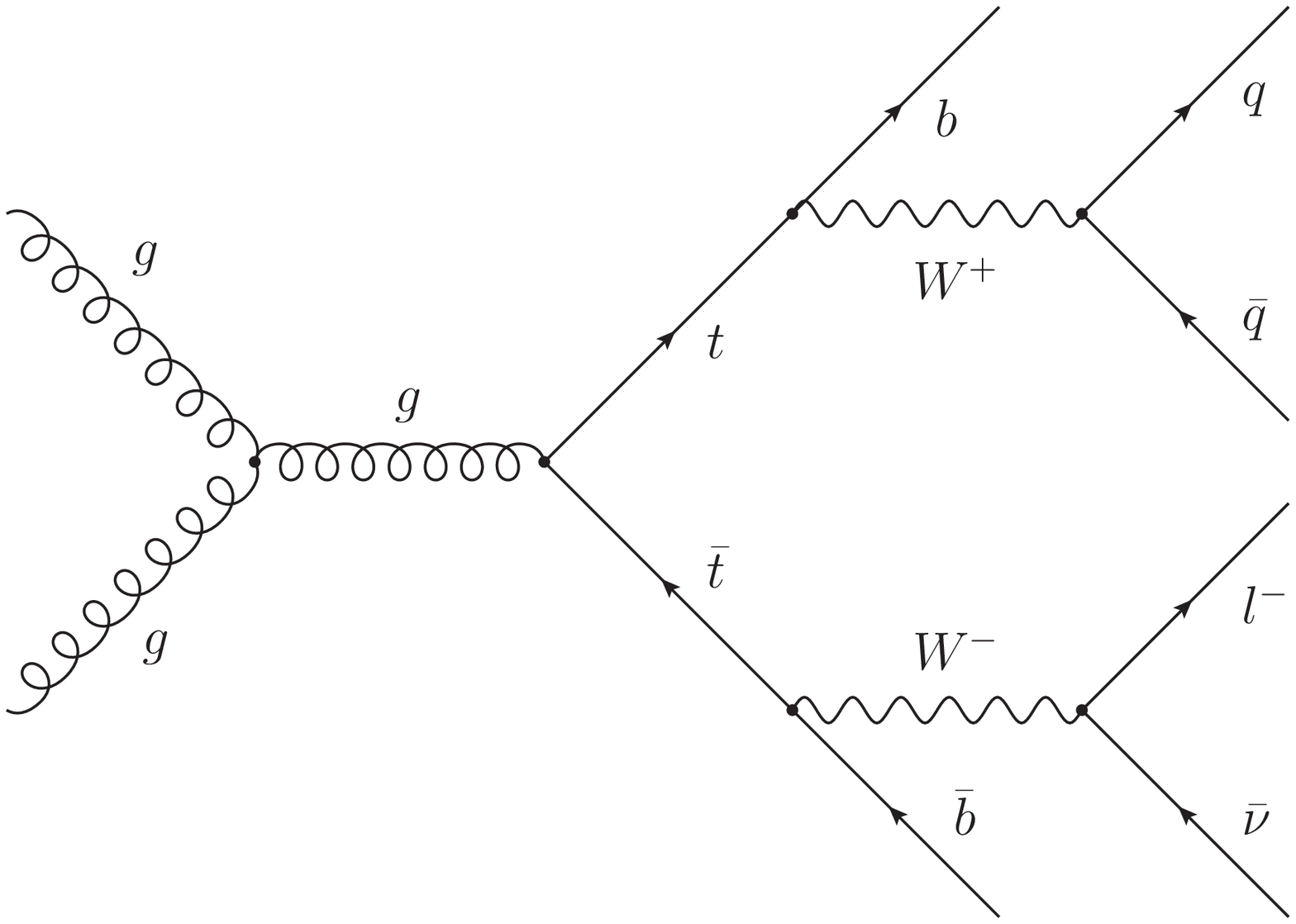,width=0.49\textwidth}{Feynman diagram of a possible fully hadronic SUSY event.\label{fullhad}}{Feynman diagram of a semileptonic $t\bar{t}$ event.\label{ttbar}}

This document is structured as follows: after the introduction, we summarize the idea of constrained kinematic fits and present a general technique based on a genetic algorithm, which simultaneously reconstructs the unmeasured particles and finds the correct jet combination (section~\ref{KinFits}). The method is tested and validated for the reconstruction of semileptonic $t\bar{t}$ events (section~\ref{ttbarfit}). Then, we show the performance of the algorithm for the example of one particular SUSY scenario (section~\ref{SusyReco}). The presented technique is finally applied to determine the masses (section~\ref{MassDetermination}), before we conclude.

\section{Event reconstruction using kinematic fits}\label{KinFits}

The reconstruction of complete events, including unmeasured particles like neutrinos or LSPs, is possible, if there are more constraints than unknowns for the problem. In case of cascade-like decays of heavy particles, the invariant masses of the intermediate particles and the momentum balance can be used as constraints. As an example, the neutrino momentum components (three unknowns) of a semileptonic $t\bar{t}$ event (see Fig.~\ref{ttbar}) can be reconstructed using the transverse momentum balance (two constraints, $\Sigma p_x = 0$ and $\Sigma p_y=0$, longitudinal boost unknown), the invariant $W$ masses (two constraints, but only the one of the leptonic branch constrains the neutrino), and the invariant top masses (two constraints, or only one in case of equal but unknown masses of the top quark). In this way it is possible to formulate an over-constrained problem, which can be solved by a constrained least square fit, taking into account the experimental uncertainties, and therefore not only reconstruct the unmeasured particles, but also improve the resolution of the measured ones.

A common technique to solve such a constrained least square problem is the method of Lagrangian multipliers (LM) \cite{Bock:1984jr}. Given $N$ measurements $y_i$ with the errors $\sigma_i$, the best values $f_i$ are to be determined, by minimizing the following expression of squared residuals
\begin{equation}
  S = \chi^2 = \sum_{i=1}^N \left({y_i - f_i\over \sigma_i}\right)^2
\end{equation}
subject to $M$ constraints of the form
\begin{equation}
  c_j(f_1, \ldots ,f_{N+P}) = 0 \quad {\rm for} \quad P {\rm~unmeasured~parameters}.
\end{equation}
The optimal values of $f_i$ yield an extremal value of the Lagrange function
\begin{equation}
  L = S + 2\cdot \sum_{j=1}^M \lambda_j\cdot c_j
\end{equation}
with the auxiliary parameters $\lambda_j$, called Lagrangian multipliers. The determination of the optimal values of $f$ and $\lambda$ are only then analytically possible, if the constraints $c_j$ are linear in $f_i$. In particular, this is not the case for invariant mass constraints which are quadratic or even more complicated, depending on the representation of the 4-momenta, as ($p_x$, $p_y$, $p_z$, $E$) or ($p_T$, $\eta$, $\phi$, $E$), where the latter is closely related to the experimental measurements and therefore less correlated. Usually, the constraints are linearized, and an iterative procedure is applied, to determine the best fitting parameters. Depending on the starting values and the number of unknown parameters, it is non-trivial to find a robust algorithm, which improves the solution at each step. Another problem is the definition of the convergence criterion, e.g.\ how well the constraints have to be fulfilled.

An alternative approach to solve a non-linear constrained least square fit is to interpret the constraints $c_j$ as additional measurements. The constraints are squared and added to the squared residual sum $S$. If the resulting expression shall be interpreted as an $\chi^2$ function, a proper normalization has to be taken into account.

In case of an invariant mass constraint of $R$ particles to a mass $m$ with a natural width $\Gamma$, the additional $\chi^2$ term $S_m$ has the form:
\begin{equation}
  S_m = {\left((\sum_{i=1}^R \mathbf{p}_i)^2 - m^2\right)^2 \over (2m\Gamma)^2}
  \label{massconstraint}
\end{equation}
Here, the approximation of normal distributed masses is assumed, to interpret the resulting expression as a $\chi^2$ term. If for different events the masses for intermediate particles at a particular position in the chains are not equal, e.g.\ because of imperfect mass degeneration of squarks, the choice of $\Gamma$ has to be done carefully (see below).

In a similar way it is possible to construct a term for the transverse momentum balance of the involved particles
\begin{equation}
  S_t = {\left(\sum_{i=1}^{\rm all~particles}p^i_{x/y}-0\right)^2 \over \sigma_t^2}. \label{sigma_t}
\end{equation}
The uncertainty of the transverse momentum balance $\sigma_t$, assumed to be equal for $x$ and $y$ direction, has to be chosen with care. In case of no initial (ISR) and final (FSR) state radiation the balance is almost perfect, due to negligible intrinsic transverse momentum in the beam. But in realistic scenarios some energy deposition outside the acceptance is breaking the balance, and $\sigma_t$ has to be given a larger value. The choice of $\Gamma$ and $\sigma_t$ has an influence on the interpretation of the fit probability.
The expression to be minimized is
\begin{equation} \label{Sprime}
  S^{\prime} = S + S_m + S_t,
\end{equation}
which can be done by various algorithms. For reasons which are discussed later, we choose an implementation of a genetic algorithm (GA).

For a given combinatorial hypothesis, the GA searches for the optimal parameter values in the following way: All momentum components of all particles (measured and unmeasured) of an event are interpreted as the genome of an individual, such that a complete event hypothesis is one individual. The fitness function, which defines if a individual is ``fitter'' than another one, is given by $S^\prime$ in Eq.~(\ref{Sprime}). Starting with a population of individuals, which can for example be created by smearing the measurements of the event\footnote{The starting values of the unmeasured particles are obtained by assuming the same flight direction as a set of measured particles, and choosing the smallest momentum, which fulfills the given mass constraint within the experimental resolution.}, the following algorithm is applied:
\begin{itemize}
  \item[(1)] Choose a few fittest individuals, according to their fitness function.
  \item[(2)] Create a new large generation of individuals by choosing randomly two parents. A new individual inherits the momentum components of each final state either from one or the other parent. Each momentum component has a small probability to mutate, which means it is to be smeared, according to a Gaussian distribution with a given width. This width can be given e.g.\ by the error of the measurement or some other carefully chosen value.
  \item[(3)] Check the convergence criterion, e.g. no change of the best individual within the last few iterations. If convergence is not reached go back to the first step. 
\end{itemize}

In summary, we traded the problems of the LM technique (linearization and convergence criterion) for computer resources, which are needed to evaluate the large amount of individuals, to find the best fitting values in the high dimensional parameter space. If the values of $\Gamma$ and $\sigma_t$ are chosen correctly, the expression $S^{\prime}$ for the best individual of each event will follow a $\chi^2$ distribution with $P-M$ (number of unmeasured parameter $-$ number of constraints) degrees of freedom.

One of the main challenges of this analysis is the suppression of the combinatorial background. The advantage of the genetic algorithm is, that the jet combinatorics can be treated simultaneously with the fit of the parameters. To achieve this, the jet combination is interpreted as an additional gene. Like the other genes, this one is inherited from either one of the two parents. For the jet assignment, the mutation is implemented by (a) simple exchange of two jets or (b) exchange of large parts of the two decay branches.

The algorithm settings, including a time dependency of the mutation, were chosen to increase the efficiency to find the correct event interpretation.

\section{Kinematic fit of $t\bar{t}$ events}\label{ttbarfit}

As a proof of principle, we performed the complete event reconstruction of semileptonic $t\bar{t}$ events (Fig.~\ref{ttbar}). Assuming the top mass is known, this results in six constraints for three unknowns, leaving overall three degrees of freedom (see above). The results of the GA are compared to a fit with the LM method \cite{KinFitter}. The studied $t\bar{t}$ sample was generated with \texttt{Pythia} \cite{Sjostrand:2006za}. For simplicity, ISR and FSR are switched off in this event generation. The true values of the final states were smeared according to typical momentum and angular resolutions for jets expected at the LHC experiments:
\begin{eqnarray}
  \frac{\sigma_{p_T}^2}{p_T^2} &=&  \left(\frac{1.0}{\sqrt{E}}\right)^2 + 0.05^2 \nonumber \\
  \sigma_\eta^2 &=& \left(\frac{0.2}{\sqrt{E}}\right)^2 + 0.02^2 \nonumber\\
  \sigma_\phi^2 &=& \left(\frac{0.3}{\sqrt{E}}\right)^2 + 0.02^2 \label{jetres}
\end{eqnarray}
Only events are fitted, in which all four jets as well as the lepton are within the acceptance of the detector, assumed here to be $p_T>30~{\rm GeV}$ and $\vert\eta\vert < 3.0$. The event selection efficiency for the semileptonic decay mode is $24\%$.

In Fig.~\ref{ttbar_ptres} and \ref{ttbar_phires} the $p_T$ and $\phi$ resolutions of the fitted neutrino are shown for the case, that the correct assignment of the jets is known. The LM converges for $98\%$ of the events for our choice of the convergence criterion ($\sum_{i=1}^M \vert c_j\vert < M\cdot 0.1$ and $\Delta S < 10^{-2}$). The GA does not reject any event by construction.

\DOUBLEFIGURE[t]{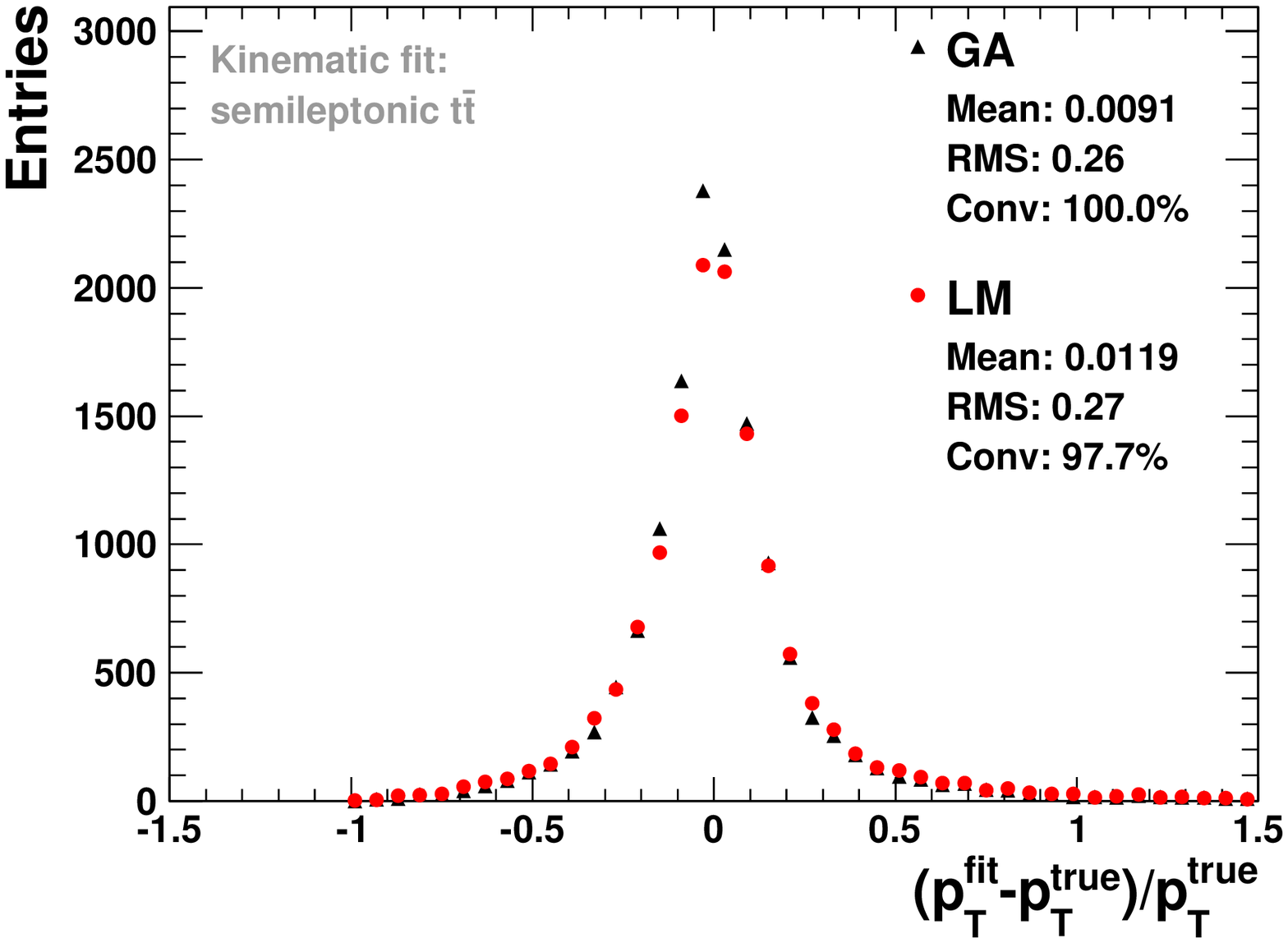,width=0.49\textwidth}{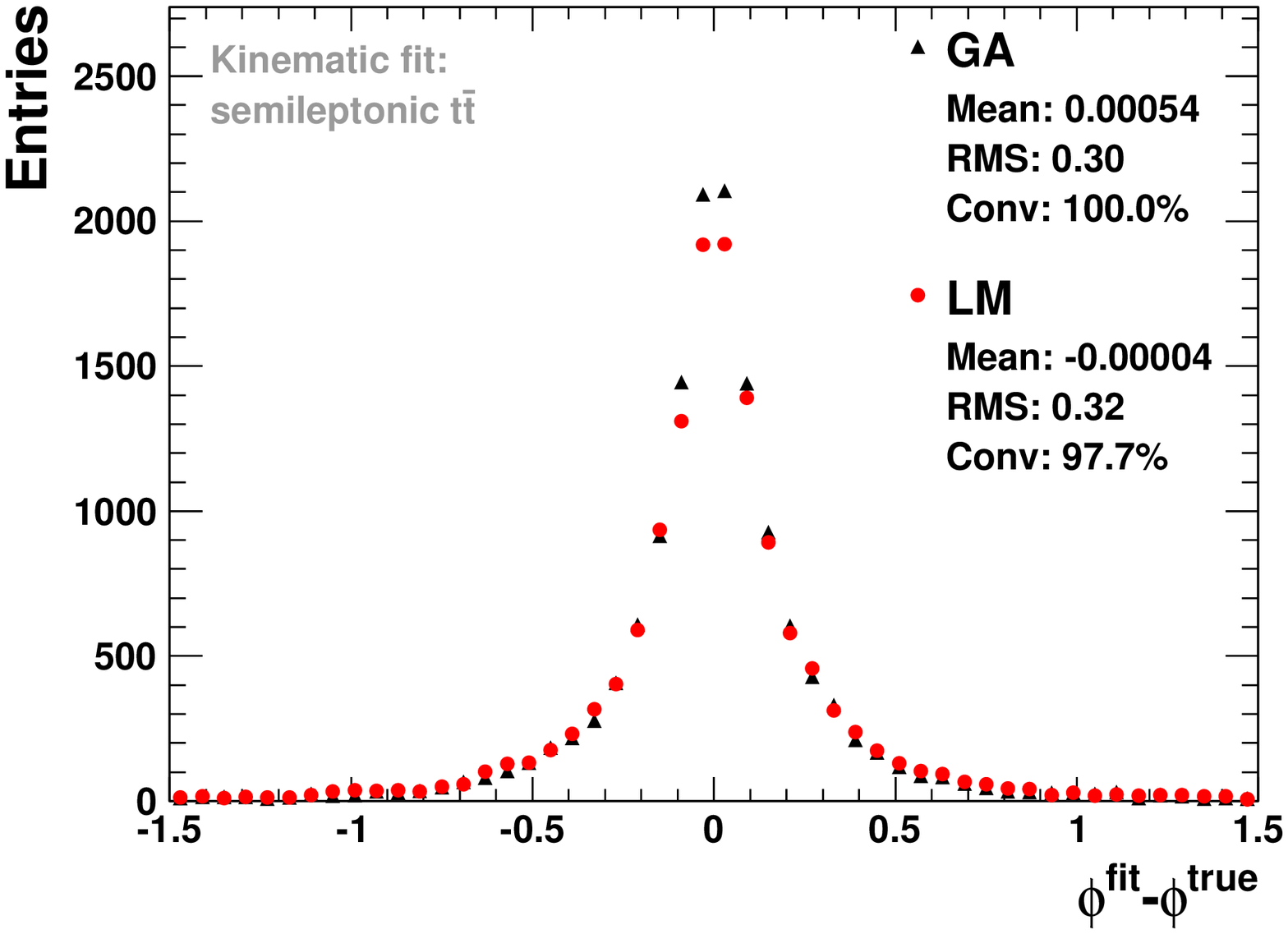,width=0.49\textwidth}{$p_T$ resolution of the fitted neutrino in semileptonic $t\bar{t}$ events.\label{ttbar_ptres}}{$\phi$ resolution of fitted neutrino in semileptonic $t\bar{t}$ events.\label{ttbar_phires}}

In Fig.~\ref{ttbar_prob} the corresponding fit probabilities are shown for both algorithms.
Both distributions are almost flat, as expected. Both algorithms show an increase at lowest fit probabilities, due to acceptance cuts and the Breit-Wigner distribution of the mass of the decaying particle.

\EPSFIGURE{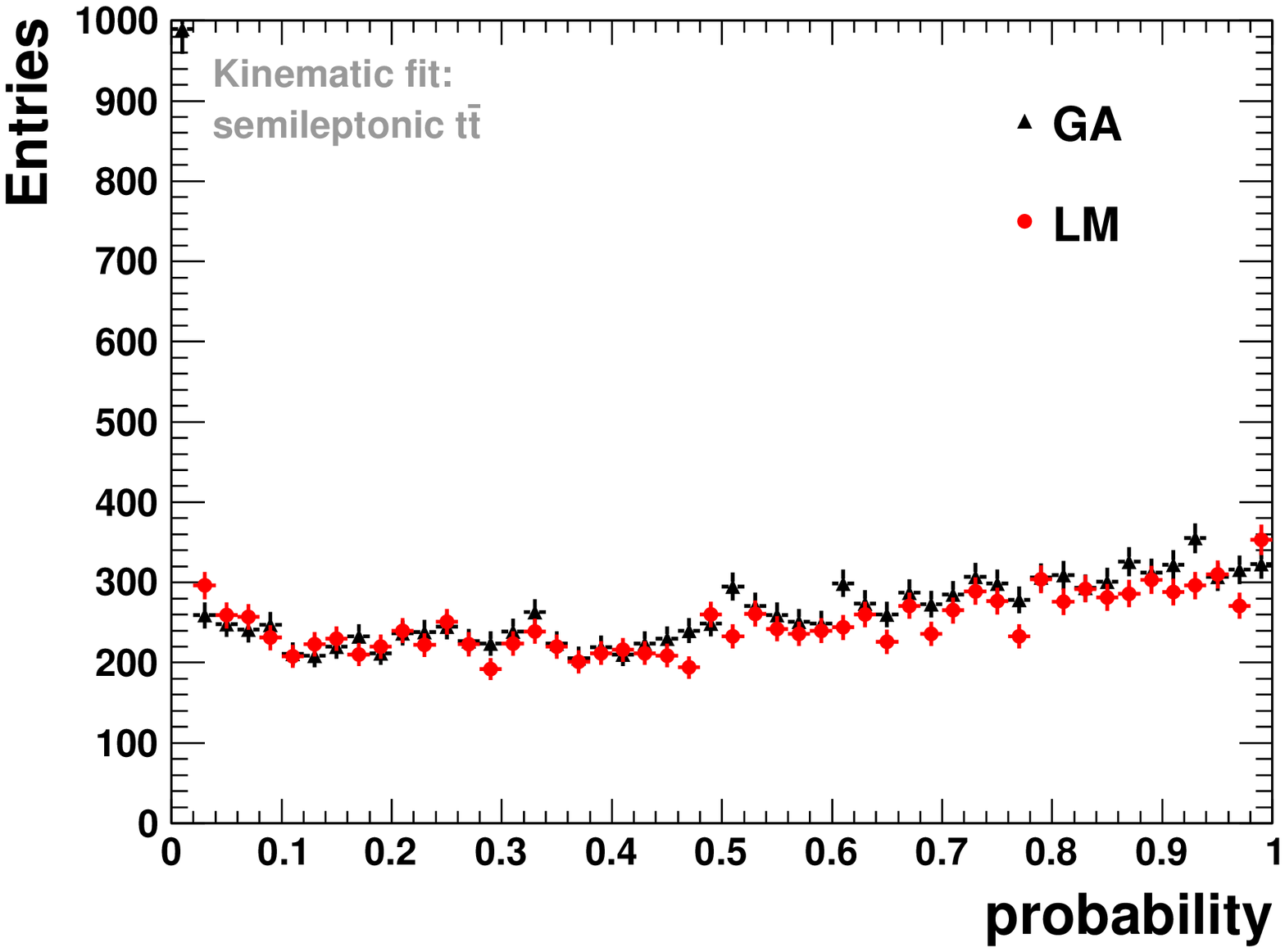,width=0.49\textwidth}{Fit probabilities for semileptonic $t\bar{t}$ events. Triangles: $\chi^2$ probability of LM fit for the correct jet assignment. Points: same as above, but for GA, interpreting $S'$ as $\chi^2$.\label{ttbar_prob}}

To study the effect of jet combinatorics, the assignment of all jets ($b$ and light quark) to the decay chains is left free to the fit. For the GA the optimal assignment is obtained during the fit process as described above. For the LM all combinations are fitted, and the one with the smallest $\chi^2$ probability is taken. For both fitting algorithms the selection of the best combinatorial hypothesis will lead to higher fit probabilities, because there is a chance for a wrong combination, to be better than the right one, whose probability is equally distributed.

In this case the GA finds for $75.6\%$ the correct combination as the best hypothesis. For the LM fit this efficiency is $72.9\%$. This difference is partially due to the convergence rate of the LM. Furthermore, in case the LM selects the correct combination, the GA finds the same hypothesis in $94.6\%$ of the cases. Both algorithms lead to the same combinatorial hypothesis in  $93.3\%$ of the events.

For this event topology, for which the number of constraints is considerably larger than the number of unknowns, the overall performance of the GA is similar, or slightly better, compared to the well established LM technique. It is therefore interesting to investigate the GA in addition to the LM for the more difficult case of SUSY event reconstruction, with more than one unmeasured particle and a factor of $\sim 10^2$ larger jet combinatorics.

\section{SUSY decay chain reconstruction}\label{SusyReco}
For the application of the kinematic fit to supersymmetric cascade decays, we choose on purpose the fully hadronic decay mode of a squark-gluino pair (see Fig. \ref{fullhad}) as the ``signal'' process, where the combinatorial problem is most severe. The gluino decays into a squark and a jet, which will be referred to as ``gluino jet''. Both squarks then decay further into a jet (``squark jet'') and either a $\tilde{\chi}_2^0$ or a $\tilde{\chi}_1^\pm$ whose masses are in many scenarios almost degenerate. The $\tilde{\chi}_2^0$ or $\tilde{\chi}_1^\pm$ decay into the $\tilde{\chi}_1^0$ (LSP) with an additional $W$, $Z$, or Higgs boson $h$ which also decays hadronically (``boson jets''). In many SUSY scenarios the fully hadronic decay modes have the largest branching ratio, and their reconstruction is most challenging due to a huge combinatorial problem. Matching the seven jets to the final states of the two decay chains, results in $1\,260$ possible combinations, not counting the irrelevant permutations of two jets of the same boson.

Since two LSPs escape the detector, resulting in six unknowns per event, at least seven constraints are needed to be over-constrained, in case of a given SUSY mass hypothesis. These are provided by the two components of the transverse momentum balance, the gluino mass, the two squark masses, and the two $\tilde{\chi}_2^0/\tilde{\chi}_1^\pm$ masses.

For a first reduction of the combinatorial background, boson jet pairs are selected by requiring their invariant mass to lie between 50 and 150~GeV, wide enough to account for the $W$, $Z$, and $h$ masses. The masses of the squarks of the first two generations are assumed to be degenerate, as expected in many SUSY scenarios. Only the left-handed squarks are considered as signal, since right-handed squarks decay dominantly directly into the LSP. 

The SUSY benchmark point chosen for this analysis has the mSUGRA parameters, masses, and properties specified in Tab.~\ref{LM5}. The mass spectrum was calculated with \texttt{SOFTSUSY} \cite{Allanach:2001kg} and the events were generated with \texttt{Pythia} \cite{Sjostrand:2006za}. 

\TABULAR{|c|c||c|c|}
{\hline
Parameter & Value & Particle & Mass [GeV] \\
\hline
$m_0$ & 230 GeV & $\tilde{q}^{L}_{ds,uc}$ & 807, 800 \\
$m_{1/2}$ & 360 GeV & $\tilde{q}^{R}_{ds,uc}$ & 775, 782 \\
$\tan\beta$ & 10 & $\tilde{q}_b^{1/2}$ & 734, 771 \\
$A_0$ & 0 GeV& $\tilde{q}_t^{1/2}$ & 599, 787 \\
sign$\mu$ & $+$ & $\tilde{g}$ & 851 \\
& & $\tilde{\chi}^0_{1,2,3,4}$ & 144, 271, 475, 490 \\
& & $\tilde{\chi}^\pm_{1,2}$ & 273, 487 \\
\hline}
{Parameters and masses of the chosen mSUGRA benchmark point. The total cross section is $\sigma^{\rm LO}_{\rm tot}\approx 1.8~$ pb for a center of mass energy$\sqrt{s}=10$ TeV.\label{LM5}}

Here, we study both the idealized case without ISR and FSR and the more realistic case with ISR and FSR in the event generation (see Tab.~\ref{selection}). Without ISR and FSR, the signal process has exactly seven jets in the final state\footnote{In our simple detector simulation we neglect the possibility of overlapping jets, which may lead to less jets in the final state. The underlying event of multi parton interaction may also lead to additional jets within the acceptance, however the number of such events is small ($<4\%$).}. From the test sample of $200\,000$ events, corresponding to a integrated luminosity of $\sim 50~{\rm fb}^{-1}$, $4.4\%$ remain after the selection cuts (no lepton, exactly seven jets with $p_{Tj}>30~{\rm GeV}$ and $\vert\eta_j\vert < 3.0$). The signal to background ratio ($S/B$) of the selected sample is $\sim 1/11$. In case of ISR and FSR the background is much increased, and furthermore only about half of the events have all the jets originating from the cascade decays within the selection cuts (see Tab.~\ref{selection}). Note that some background events have a rather signal-like decay topology, for example via third generation squarks or via heavier neutralinos or charginos.

\TABULAR{|c|cc|cc|}
{\hline
& \multicolumn{2}{|c|}{without ISR \& FSR} & \multicolumn{2}{|c|}{with ISR \& FSR}\\
& Bg & Sig & Bg & Sig \\
\hline
selection efficiency & 4.2\% & 29\% & 6.9\% & 30\%\\
$S/B$ & \multicolumn{2}{|c|}{1/10.9} & \multicolumn{2}{|c|}{1/16.4} \\
$S/B$ (complete) & \multicolumn{2}{|c|}{1/11.3} & \multicolumn{2}{|c|}{1/33.4} \\
\hline}
{Selection efficiencies and $S/B$; ``complete'' means that all jets from the cascade decays are within the selection cuts. In not ``complete'' events at least one jet of the cascade decay is outside the acceptance cuts, but instead an ISR, FSR or underlying event jet is selected.\label{selection}}

In a first step, we aim for the complete reconstruction of SUSY events for a fixed set of SUSY masses. Therefore, in this section the masses are set to their true values, while in section \ref{MassDetermination} a scan is performed to determine the masses.

\subsection{Kinematic fit}

In the following, we present the fit results including ISR and FSR in the event simulation. Due to ISR, the transverse momentum balance of the $\tilde{q}\tilde{g}$ decay products has a width of $\sigma_t \approx 40~{\rm GeV}$ (see Eq.~\ref{sigma_t}). To compensate for this, the measured jets outside the acceptance are added to the momentum balance, improving the above number to $\sigma_t\approx 15~{\rm GeV}$ for the fit. For the SUSY mass constraints, the natural width as well as effects from imperfect mass degeneration are taken into account ($\Gamma_{\tilde{g}}=10$~GeV, $\Gamma_{\tilde{q}}=15$~GeV and $\Gamma_{\tilde{\chi}_2^0/\tilde{\chi}_1^\pm}=5$~GeV).

In order to compare the performance of the GA and the LM, we study the $p_T$ and $\phi$ resolution of the fitted neutralinos (see Fig.~\ref{LM5_ptres} and \ref{LM5_phires}). The fraction of events for which the GA finds the correct jet assignment as the best hypothesis is $42\%$ which has to be compared with $1/1\,260$ in the case of random combinatorics. In $\sim 10\%$ of the remaining events, the best hypothesis is given by the exchange of the decay branches from the two squark decays. In this case all mass constraints are also fulfilled, except for the gluino mass constraint, which depends on the relativly soft gluino jet\footnote{Due to the relative small mass gap between the gluino and the squarks, in comparison to the gap between squarks and $\tilde{\chi}_2^0/\tilde{\chi}_1^\pm$, the squark jet is on average harder than the gluino jet.} with its larger relative uncertainty (see Eq.~\ref{jetres}). In the remaining cases wrong hypotheses are selected due to combinations with an accidentally small $S'$.

\DOUBLEFIGURE[t]{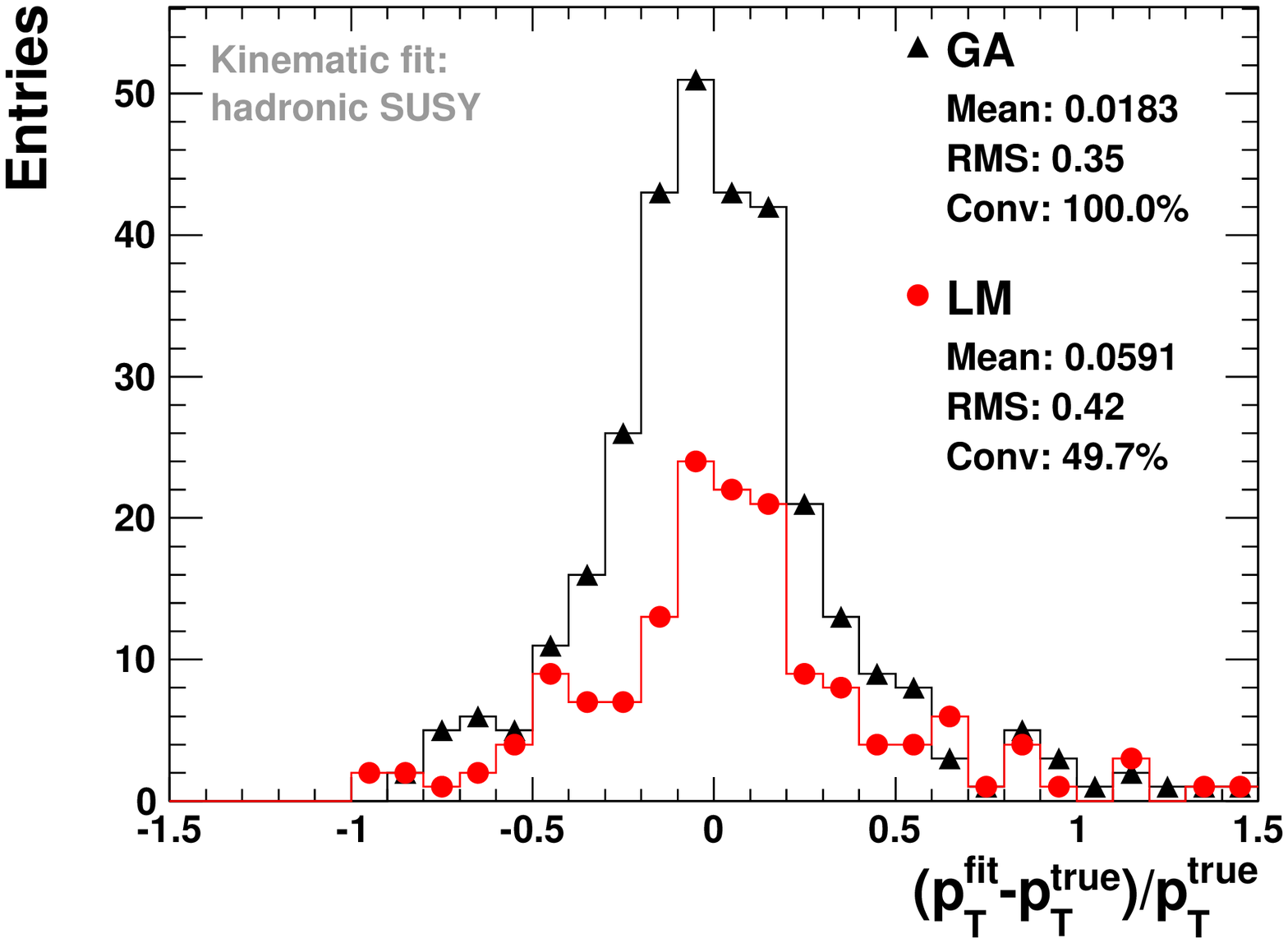,width=0.49\textwidth}{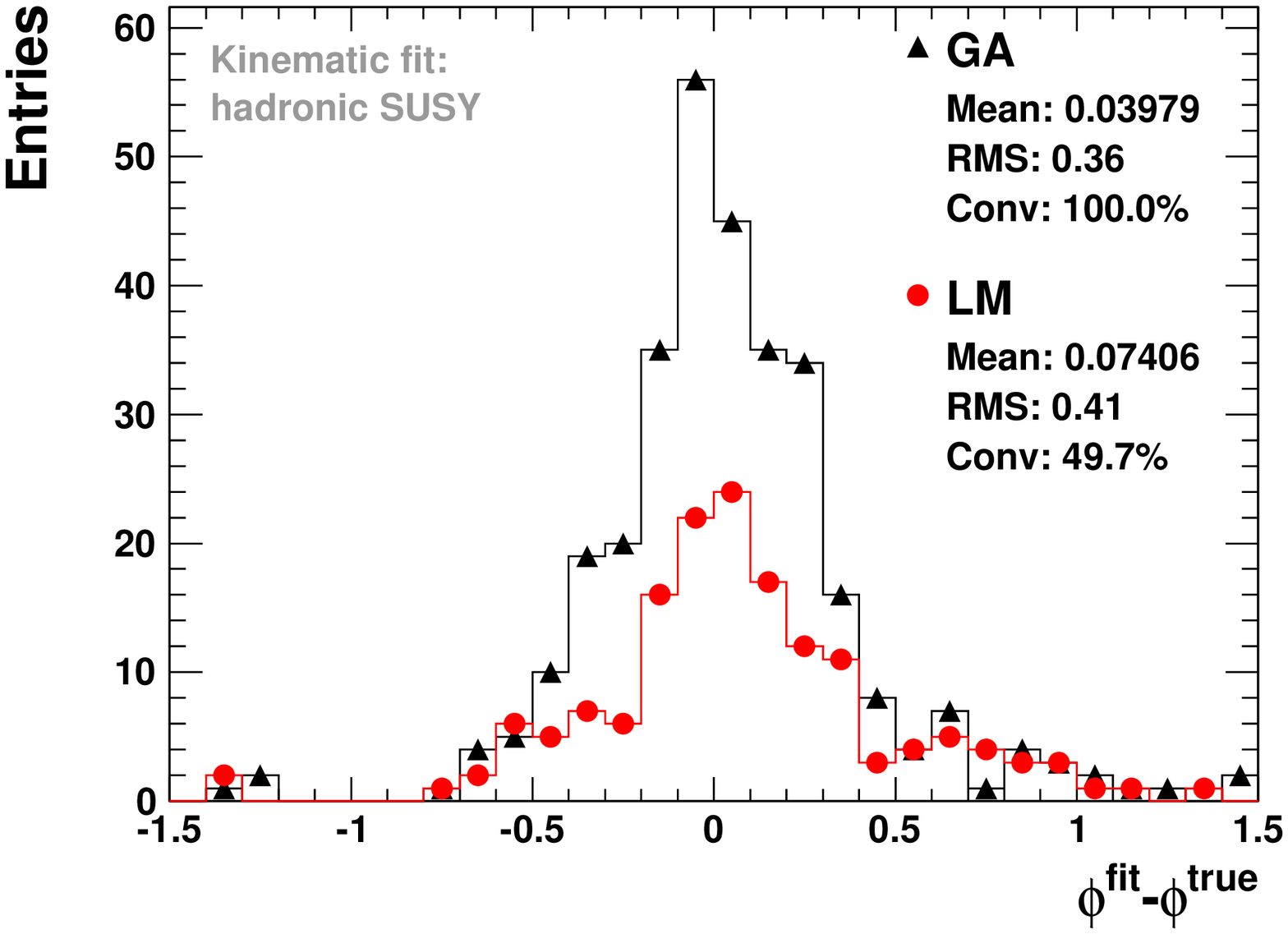,width=0.49\textwidth}{$p_T$ resolution of fitted neutralinos. Only events are shown where the best hypothesis has the correct jet assignment.\label{LM5_ptres}}{$\phi$ resolution of fitted neutralinos. Only events are shown where the best hypothesis has the correct jet assignment.\label{LM5_phires}}

The definition of a convergence criterion for the LM is difficult. Choosing a setting which yields a small bias and resolutions in $p_T$ and $\phi$ similar to the GA, we obtain the same ``purity'' ($41\%$) of correct jet assignments, however at the cost of a much reduced convergence rate of $\sim 50\%$. Due to its higher rate, the GA is the algorithm of choice for the further studies.

\EPSFIGURE{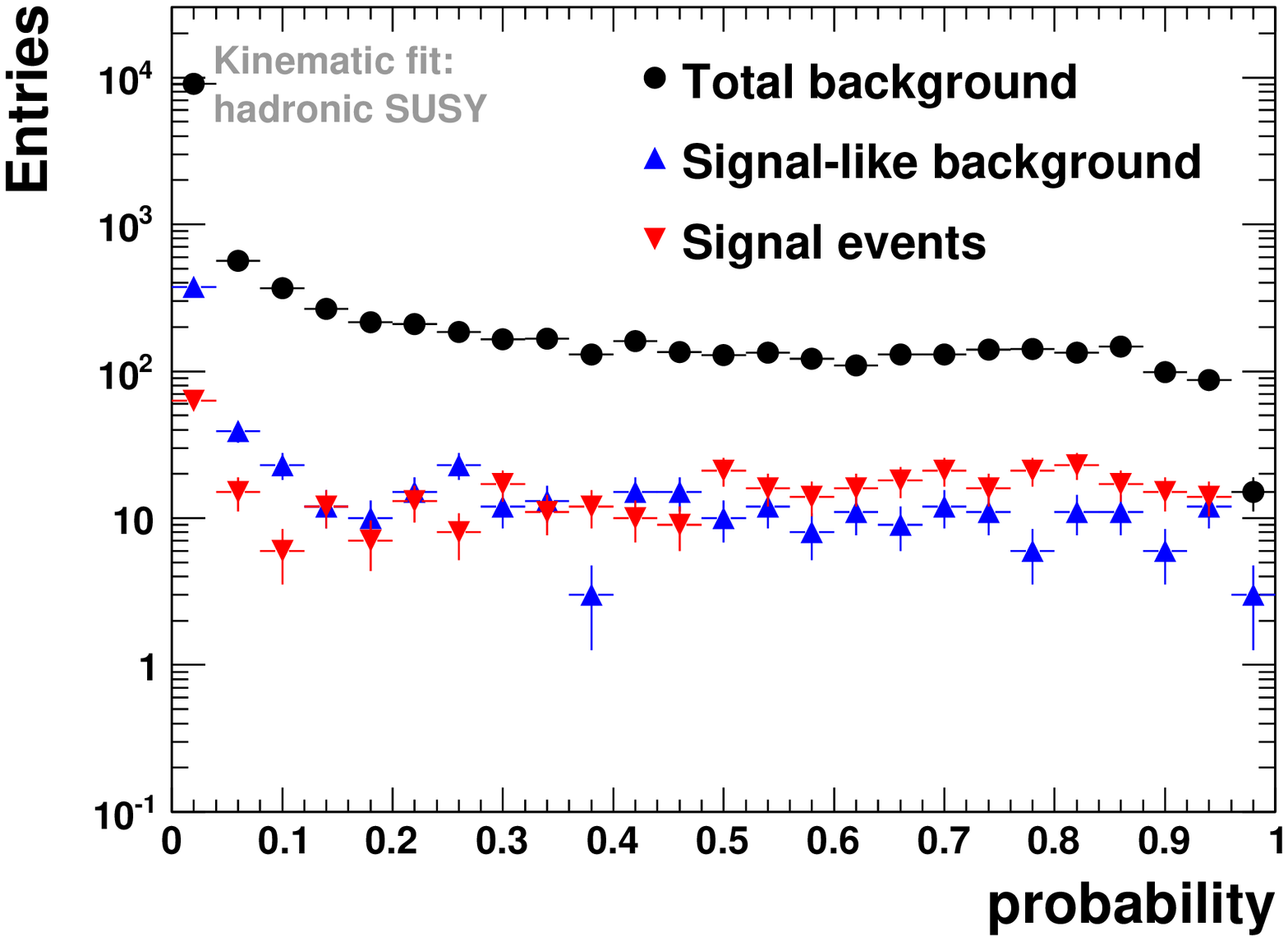,width=0.49\textwidth}{Fit probability for best event hypotheses using the GA.\label{LM5_prob}}

In Fig.~\ref{LM5_prob} the fit probability $p$ for the GA is shown. Above $p>0.1$ the distribution looks reasonably flat for the signal, as expected. Possible differences to a flat distribution are the jet combinatorics, which leads to a shift towards higher probabilities, and under- or overestimation of uncertainties. In addition, the distributions are shown for the total background, as well as for ``signal-like'' background. The latter consists of either signal events where one jet of the cascades lies outside the acceptance cuts and instead an ISR or FSR jet is found, or decays with the correct decay topology but involving squarks of the third generation. The number of such events is of the same order as the number of signal events. They also show a rather flat probability distribution for values beyond $0.1$. In contrast, the background events show a clear deviation from the uniform distribution with a strong increase towards lower values. As a consequence, a cut on $p$ can be used to improve the signal to background ratio. In this example, the total $S/B$ of $1/33$ is improved to $1/11$ ($1/8$) for a cut at 0.1 (0.3) at the cost of reducing the signal efficiency by 79\% (68\%). It is noticed, that beside the rather simple selection cuts no further reduction of the SUSY background has been attempted here.

\subsection{Angular distributions}\label{angular}

So far, we have only utilized invariant mass and transverse momentum balance constraints. Further information of the event can be used, like for example decay angles. In Fig.~\ref{LM5_cosThetaCha} and \ref{LM5_cosThetaSq} the $\cos\theta^\ast$ distribution of the decay products in the rest frame of the decaying $\tilde{\chi}^\pm_1/\tilde{\chi}^0_2$ and squark are shown, where $\theta^\ast$ is the fitted angle with respect to the flight direction of the decaying particle. In case of a spin-0 or unpolarized particle, the expected distribution is uniform, as can be seen for the correct jet combinations. The acceptance cuts on the jets influence mostly the distribution for the $\tilde{\chi}^\pm_1/\tilde{\chi}^0_2$ decay: The lower $p_T$ cut prefers bosons decaying into the flight direction of the SUSY particle, since then the two jets have on average higher $p_T$. This leads to a slight increase in the $\cos\theta^\ast$ distribution towards $+1$. For the background consisting of other decay modes the distribution is very different.

\DOUBLEFIGURE[t]{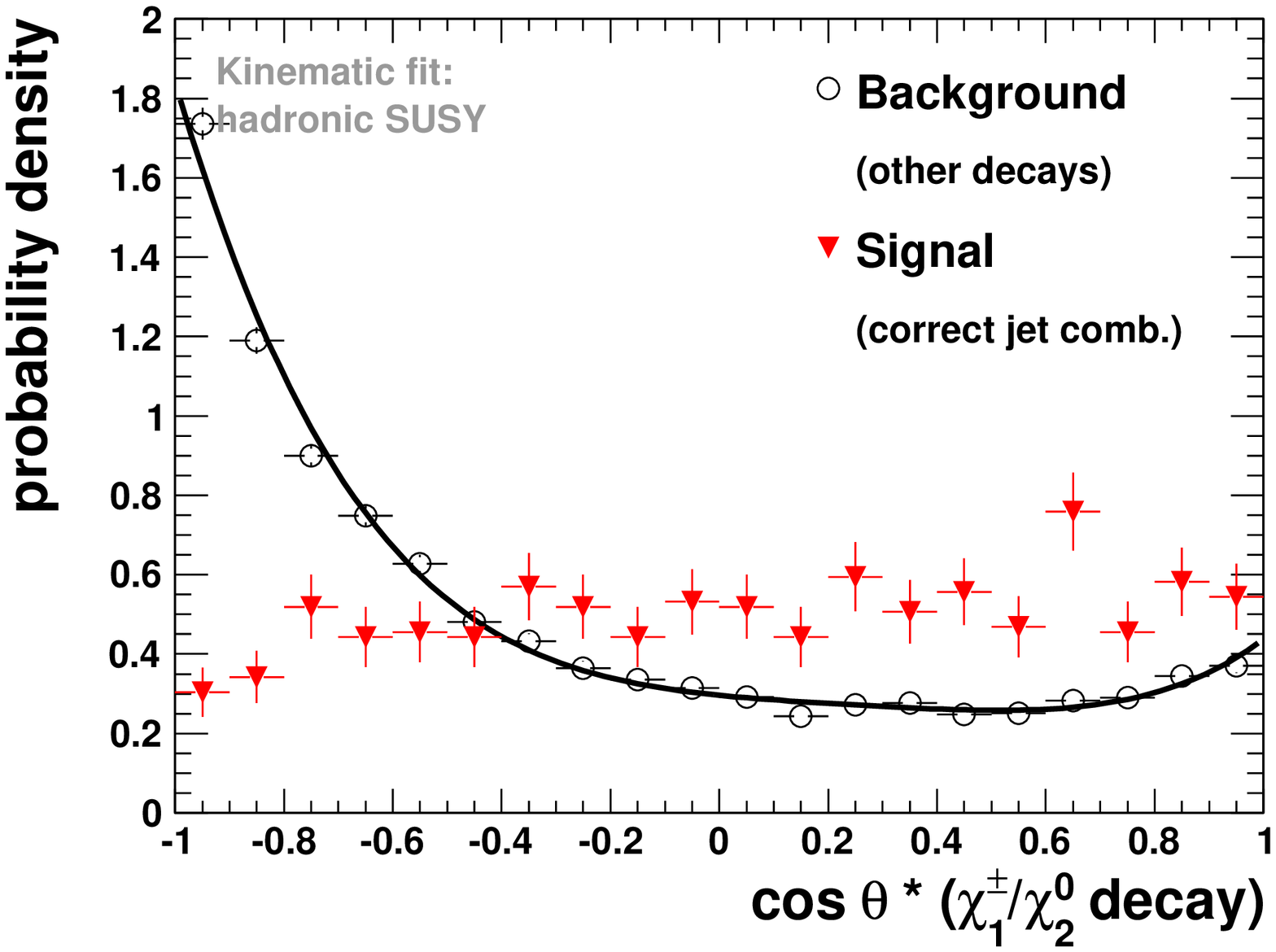,width=0.49\textwidth}{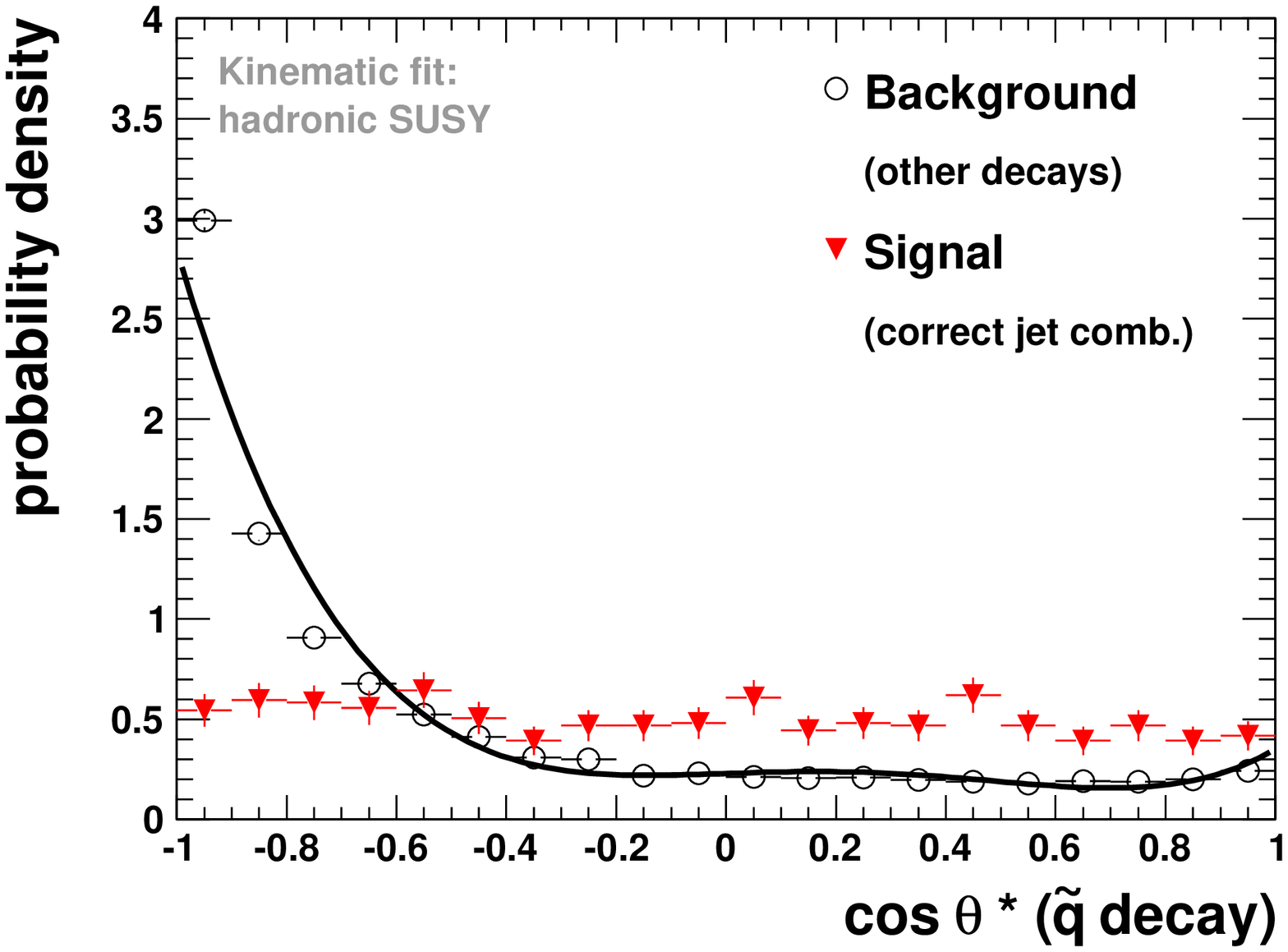,width=0.49\textwidth}{Angular distribution of the boson with respect to the flight direction of the $\tilde{\chi}^\pm_1/\tilde{\chi}^0_2$ in the rest frame of the decay, for the GA fit.\label{LM5_cosThetaCha}}{Angular distribution of the squark jet with respect to the flight direction of the squark in the rest frame of the decay, for the GA fit.\label{LM5_cosThetaSq}}

The normalized distribution of the background is parametrized by a polynomial and then used as a probability density $f$ for a likelihood ratio. The signal distribution is assumed to be flat. The probability density to be signal-like of an event hypothesis with a set of $N$ decay angles $\cos\theta^\ast_i$ can then be expressed by:
\begin{equation}
  LR = {1\over C_{\rm norm}}\prod_{i=1}^N {f_{\rm sig}(\cos\theta^\ast_i) \over f_{\rm sig}(\cos\theta^\ast_i) + f_{\rm bg}(\cos\theta^\ast_i)}
\end{equation}
The product $L$
\begin{equation}
  L = p \cdot LR \label{L}
\end{equation}
is used to define the best hypothesis, including angular information of the event. In this case $L$ helps for a further suppression of the SUSY and SM background. In our example a cut on $L$ with a signal efficiency of 77\% leads to $S/B$ of $1/9$, while a cut on $p$ at 0.1 with a similar efficiency of 78\% yields $1/11$. For the reduction of the combinatorial background, the performance of $L$ and $p$ are similar. It should be pointed out, that the probability distribution of the background depends on the model, which was used for this study. Further studies have to be made to reduce this model dependence.

\section{Constraining SUSY masses}\label{MassDetermination}

In the following the mass reconstruction is studied using the GA including full combinatorics but ignoring other SUSY processes. For this a scan of $m(\tilde{q})$ and $m(\tilde{\chi}^\pm_1/\tilde{\chi}^0_2)$ is performed within the allowed kinematic range, fixing $m(\tilde{g})$ and $m(\tilde{\chi}^0_1)$ to their true values. Although two masses cannot be constrained by a single event, a set of events leads to an over constrained system. The true mass hypothesis is expected to result on average in a better fit probability. There are various measures to quantify this behavior:

\begin{itemize}
   \item Number of converging events $N_{\rm converged}$: In the case of the GA ``converging'' means that the fit probability is above a given threshold $p_{\rm cut}$. In the following we choose $p_{\rm cut}=0.01$.
   \item Average fit probability of converging events: This measure has the disadvantage that single events with a good fit probability for a wrong mass hypothesis can suggest "good" regions in the mass space, although only a few events are converging at all.
   \item Fractional event count $N_{\rm frac}$: This is defined as the sum of the fit probabilities of all events. This quantity has only a small dependence on the number of converged events, since not converged events do not contribute due to their small probability. However, a statistical interpretation is difficult.
   \item Normalized likelihood of all events: The product $\mathcal{P}$ of the fit probabilities $p_i$ for all events is a natural definition of the overall likelihood of a given mass hypothesis. Unfortunately, the number of converging events is not constant for all mass hypotheses. The likelihood can be normalized to the number of events as follows:
   \begin{equation}
     \mathcal{P}_{\rm norm} = \left(\prod_{i=1}^{N_{\rm converged}} p_i\right)^{1\over N_{\rm converged}}= \exp\left({1\over N_{\rm converged}}\sum_{i=1}^{N_{\rm converged}}\log p_i\right)
   \end{equation}
   In regions where the number of converging events is small, $\mathcal{P}_{\rm norm}$ suffers from the same fluctuations as described for the average fit probability.
   \item Regularized likelihood: To avoid the problems of normalized quantities, each not converged event can be assigned a small probability $p_{\rm cut}$. In this case, the same number of events are used for each mass hypothesis, and $\mathcal{P}$ does not have to be normalized. The logarithm of $\mathcal{P}$ has numerical advantages and can be interpreted statistically.
   \begin{equation}
      \log\mathcal{P} = \sum_{i=1}^{N_{\rm tot}} \log{{\rm max}(p_i, p_{\rm cut})}
   \end{equation}
   Note, that the value of $p_{\rm cut}$ is not critical since the statistical interpretation depends on $\Delta\log\mathcal{P}$ relative to the maximum, and close to the maximum variations are negligible.
\end{itemize}

In Fig.~\ref{LM5_scan_sig_right} (\ref{LM5_scan_sig_wrong}) the fractional event count $N_{\rm frac}$ is shown in the $\tilde{\chi}^\pm_1/\tilde{\chi}^0_2-\tilde{q}$ mass plane for the events where the fit finds the correct (wrong) jet assignment. For the events where the correct hypothesis was found, the true masses lie within $10~{\rm GeV}$ around the maximum, in contrast to the wrong combinations which prefer a larger $\tilde{\chi}^\pm_1/\tilde{\chi}^0_2$ mass. We attribute this to the larger combinatorics for the $\tilde{\chi}^\pm_1/\tilde{\chi}^0_2$ at heavier mass hypotheses, as well as to jet threshold effects at low mass hypotheses. In principle, if all the uncertainties are known, this bias can be extracted from Monte-Carlo simulations.

\DOUBLEFIGURE[t]{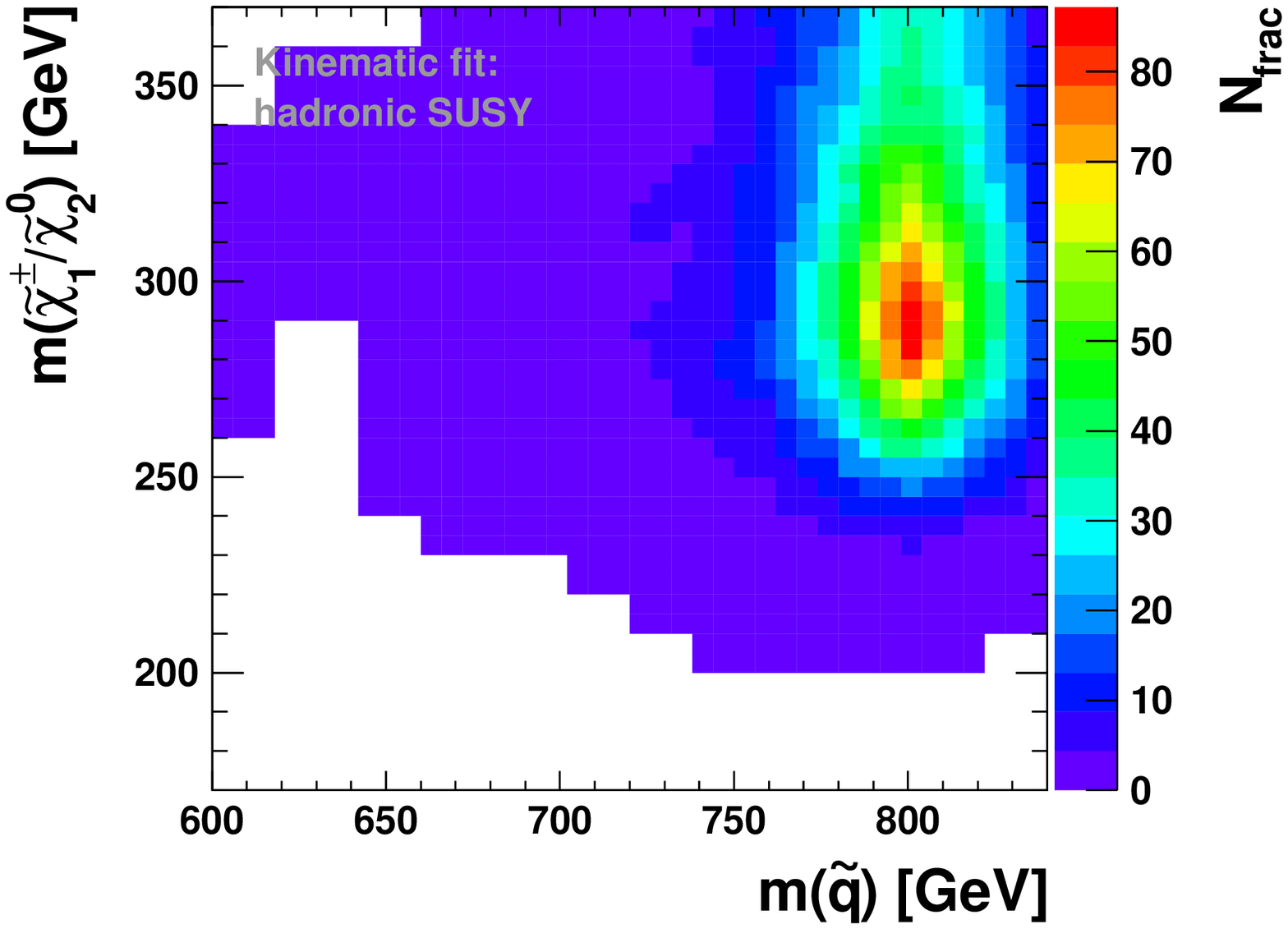,width=0.49\textwidth}{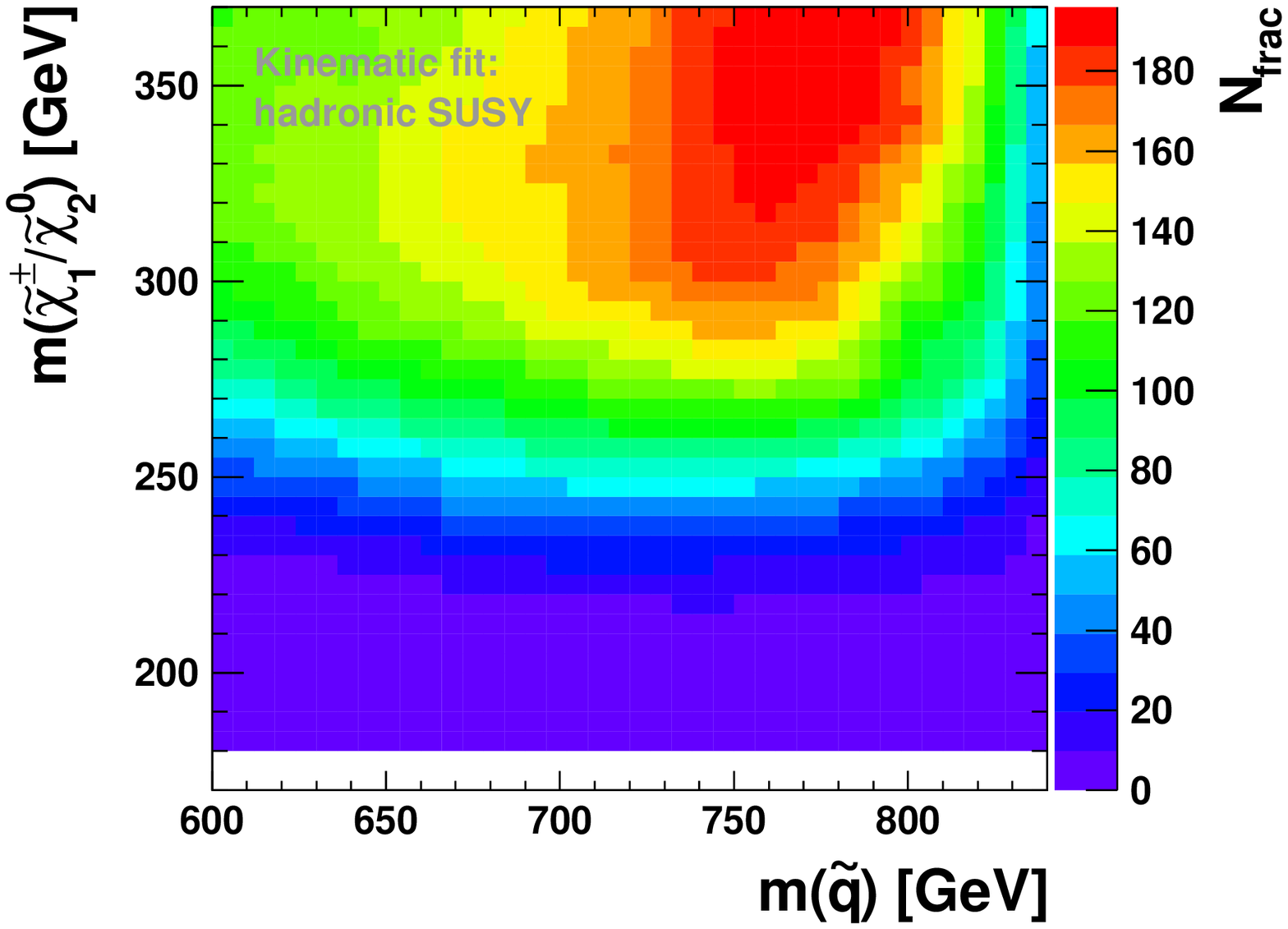,width=0.49\textwidth}{$N_{\rm frac}$ of events for which the GA finds the correct jet assignment. ISR and FSR are included.\label{LM5_scan_sig_right}}{$N_{\rm frac}$ of events for which the GA finds the wrong jet assignment. ISR and FSR are included.\label{LM5_scan_sig_wrong}}

\EPSFIGURE{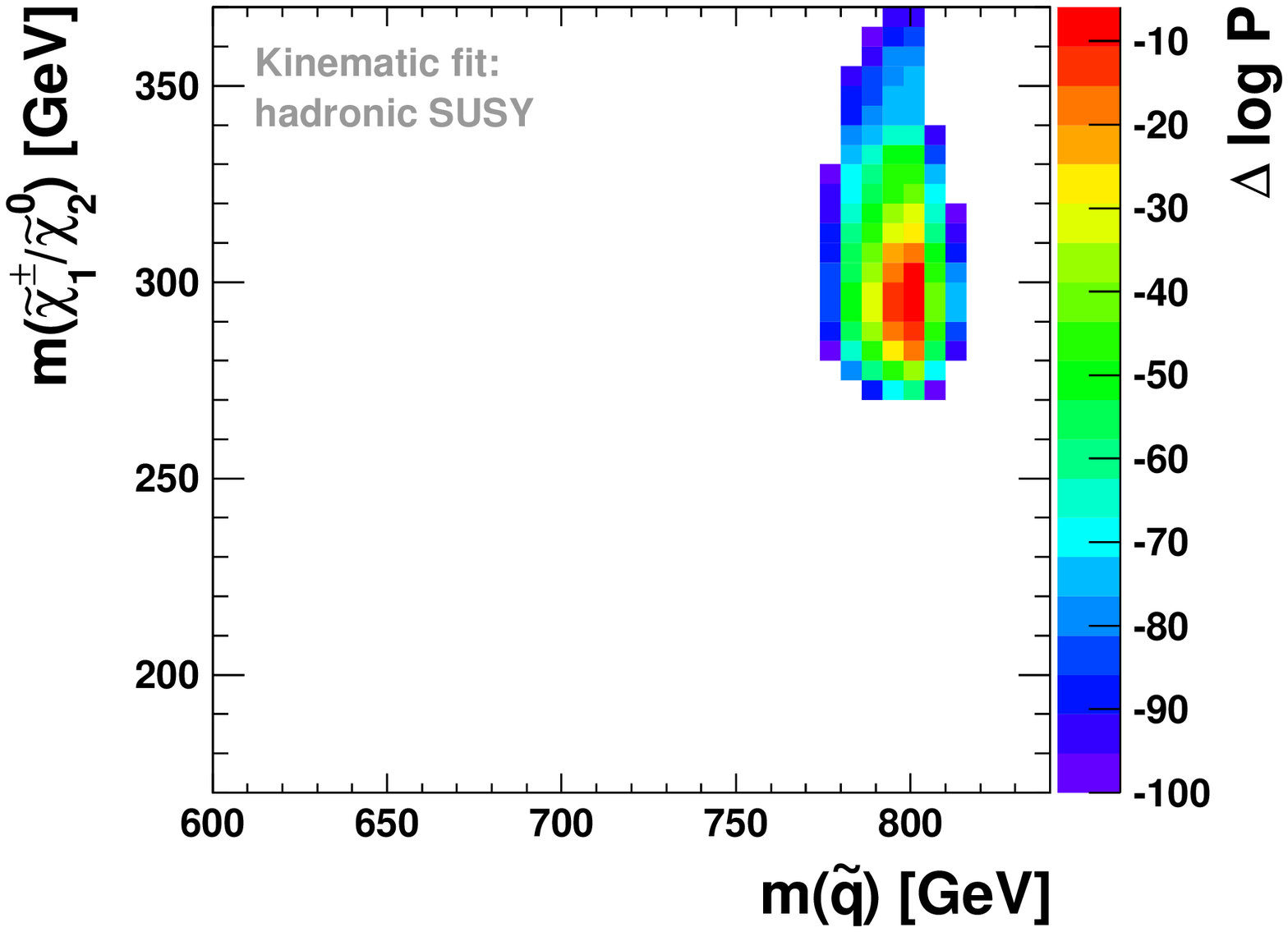,width=0.49\textwidth}{$\Delta\log\mathcal{P}$ of signal events, including ISR and FSR, and without combinatorial background.\label{LM5_noIFSR_scan_sig_nocomb}}

Fig.~\ref{LM5_noIFSR_scan_sig_nocomb} shows $\Delta\log\mathcal{P}$ without combinatorial background. As for the case of $N_{\rm frac}$ the true masses are found with a precision similar to the width $\Gamma$ of the mass constraints. If the combinatorial background is included this picture changes as shown in Fig.~\ref{LM5_scan_sig}. A clear correspondence between the maximum of $\Delta\log\mathcal{P}$ and the true squark mass of $800~{\rm GeV}$ is visible. However, there is a systematic shift towards larger $\tilde{\chi}^\pm_1/\tilde{\chi}^0_2$ masses, with respect to the true value of $270~{\rm GeV}$. In Fig.~\ref{LM5_noIFSR_scan_sig} $\Delta\log\mathcal{P}$ is shown for a sample without ISR and FSR. In this case the momentum balance is a much stronger constraint, since additional jets outside the acceptance cuts are not present. As a consequence the GA finds in 48\% of the events the correct jet assignment, compared to 42\% with ISR and FSR included, and the systematic shift towards larger $\tilde{\chi}^\pm_1/\tilde{\chi}^0_2$ masses is significantly reduced.

\DOUBLEFIGURE{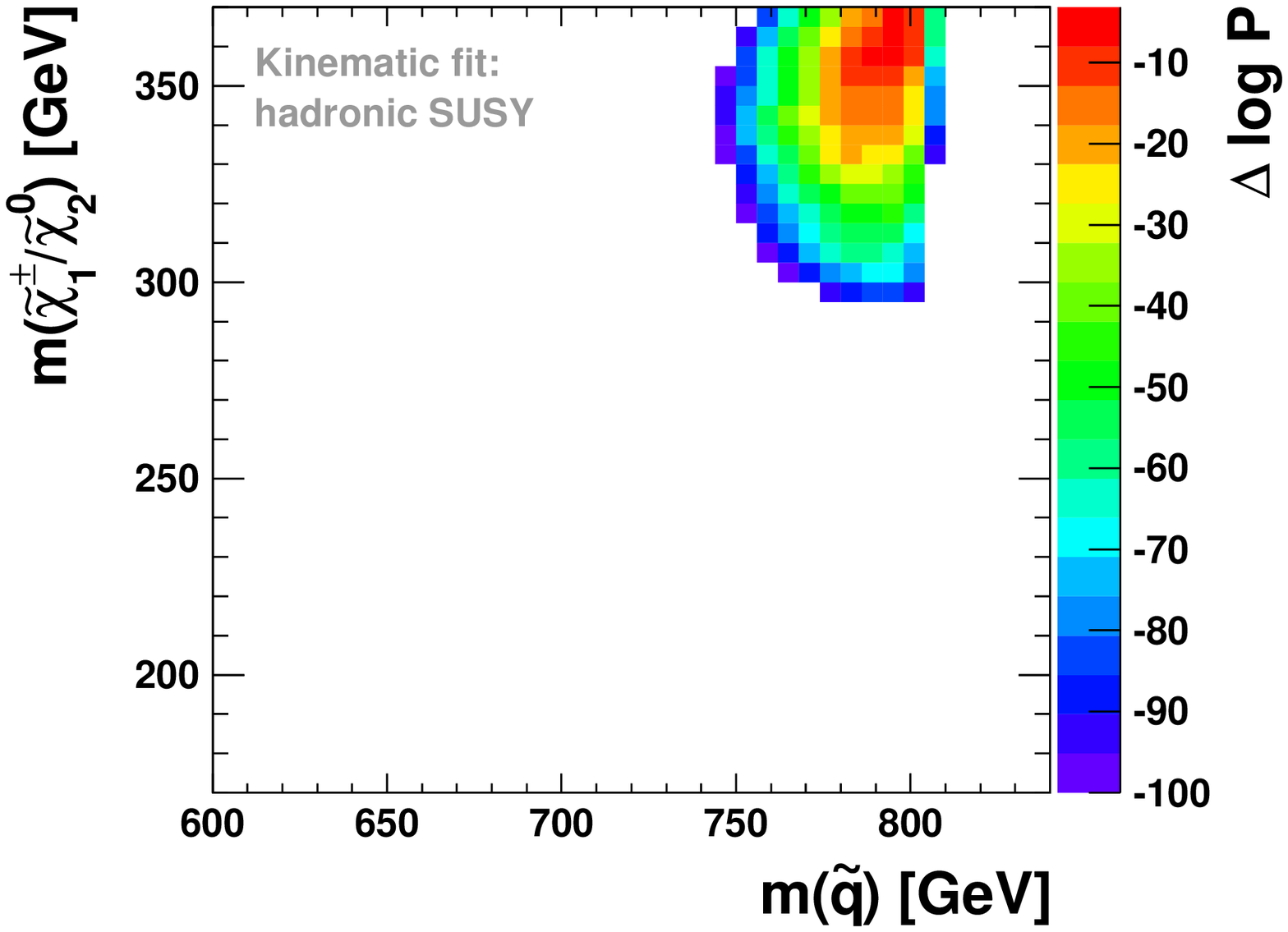,width=0.49\textwidth}{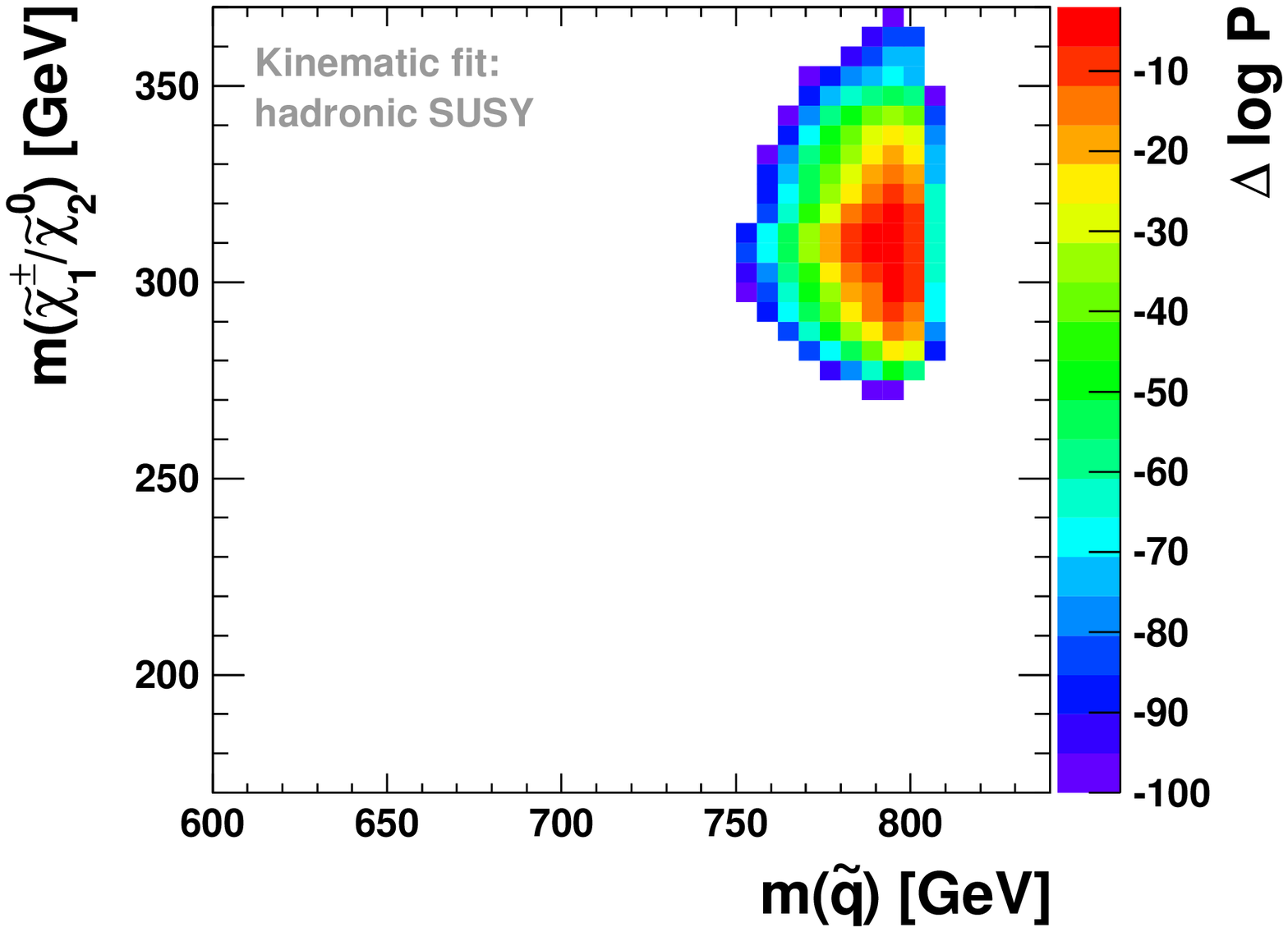,width=0.49\textwidth}{$\Delta\log\mathcal{P}$ of signal events. The true masses are $m(\tilde{\chi}^\pm_1/\tilde{\chi}^0_2)\approx {\rm 270~GeV}$ and $m(\tilde{q})\approx {\rm 800~GeV}$. ISR and FSR are included.\label{LM5_scan_sig}}{Same quantity as on the left hand side, but no ISR and FSR present.\label{LM5_noIFSR_scan_sig}}

If the SUSY background is included, the interpretation will be more difficult, depending on the signal to background ratio. A combination with other observables, like event rates or mass edges from other final states, is statistically possible.

\section{Conclusion}\label{Conclusion}

We have presented a novel approach to perform constrained kinematic fits based on a genetic algorithm, and applied it to the complete reconstruction of supersymmetric decay chains of $\tilde{q}\tilde{g}$ production with two unmeasured particles. The fully hadronic decay mode with seven jets in the final state was chosen, since it has in many SUSY scenarios a large branching fraction, and the large jet multiplicity leads to a huge combinatorics, which is a viable test for the developed fitting technique. In comparison with a conventional kinematic fitting based on the method of Lagrangian multipliers, we observe the genetic algorithm to converge for a factor of two more events and to find in $\sim 45\%$ of the events the correct jet assignment. In addition, decay angles were used to improve the rejection of other SUSY processes. In this way more information than only invariant mass constraints were used.

A SUSY mass scan was performed, and a regularized likelihood was used to derive a probability map of the SUSY mass space. We note that squark masses are reconstructed quite well, if samples with only signal events are used. The $\tilde{\chi}^\pm_1/\tilde{\chi}^0_2$ masses are reconstructed with a systematic shift towards larger values. This effect is mainly due to events where the fit finds a wrong jet assignment, but can be corrected using Monte-Carlo simulations.

For samples including the full SUSY background further selection cuts to improve the signal to background ratio have to be studied. The application of this method to leptonic decay modes will be of particular interest, due to reduced combinatorics, less SUSY and SM backgrounds, as well as more accurate final state measurements.

\end{document}